\documentclass[aps,prd,twocolumn,amsmath,amssymb]{revtex4-2}
\usepackage[utf8]{inputenc} 
\usepackage[T1]{fontenc}    
\usepackage{lmodern}
\usepackage{hyperref}       
\usepackage{amsfonts}       
\usepackage{amsmath,amsthm,amssymb} 
\usepackage{bm}
\usepackage{bbm}
\usepackage{dcolumn}
\usepackage{nicefrac}       

\usepackage{graphicx}
\usepackage{xcolor, etoolbox}
\usepackage{mathtools}
\usepackage{mathrsfs}
\usepackage{microtype}      
\usepackage{braket}      
\usepackage{enumitem}
\usepackage{xspace}
\usepackage{dsfont}
\usepackage{float}

\hypersetup{
    colorlinks,
    linkcolor=blue,
    citecolor=blue,
    urlcolor=blue,
    breaklinks=true 
}

\graphicspath{{./figures/}} 

\usepackage[normalem]{ulem} 

\begin{document}

~\hfill {IPARCOS-UCM-23-019}

\title{Operational realization of quantum vacuum ambiguities}

\author{\'Alvaro \'Alvarez-Dom\'inguez}
	\email{alvalv04@ucm.es}
    \affiliation{Departamento de F\'isica Te\'orica and IPARCOS,
Universidad Complutense de Madrid, Plaza de las Ciencias 1, 28040 Madrid, Spain}  

\author{Jos\'e A.R. Cembranos}
	\email{cembra@ucm.es}
    \affiliation{Departamento de F\'isica Te\'orica and IPARCOS,
Universidad Complutense de Madrid, Plaza de las Ciencias 1, 28040 Madrid, Spain}  

\author{Luis J. Garay}
	\email{luisj.garay@ucm.es}
    \affiliation{Departamento de F\'isica Te\'orica and IPARCOS,
Universidad Complutense de Madrid, Plaza de las Ciencias 1, 28040 Madrid, Spain}  

\author{Mercedes Mart\'in-Benito}
	\email{m.martin.benito@ucm.es}
    \affiliation{Departamento de F\'isica Te\'orica and IPARCOS,
Universidad Complutense de Madrid, Plaza de las Ciencias 1, 28040 Madrid, Spain} 

\author{\'Alvaro Parra-López}
	\email{alvaparr@ucm.es}
    \affiliation{Departamento de F\'isica Te\'orica and IPARCOS,
Universidad Complutense de Madrid, Plaza de las Ciencias 1, 28040 Madrid, Spain} 

\author{Jose M. Sánchez Velázquez}
	\email{jm.sanchez.velazquez@csic.es}
    \affiliation{Instituto de F\'isica Te\'orica UAM/CSIC, c/ Nicol\'as Cabrera 13-15,  Cantoblanco, 28049 Madrid, Spain}

\date{\today}

\begin{abstract}
We provide a reinterpretation of the quantum vacuum ambiguities that one encounters when studying particle creation phenomena due to an external and time-dependent agent. We propose a measurement-motivated understanding: Each way of measuring the number of created particles selects a particular vacuum. This point of view gives a clear and physical meaning to the time evolution of the number of particles produced by the agent as the counts in a specific detector and, at the same time relates commonly used quantization prescriptions to particular measurement setups.
\end{abstract}
\maketitle


\section{Introduction}

When quantizing a matter field in the presence of a classical, external, and time-dependent agent, there arises the phenomenon of particle production. Examples of this are gravitational particle creation caused by a time-dependent geometry \cite{Parker1969, Ford1987, Weinberg2008} or pair production in the Schwinger effect due to an external electric field~\cite{Sauter1931,Schwinger1951}.

These phenomena are usually studied within the framework of quantum field theory in curved spacetimes \cite{Birrell1982, Mukhanov2007, Calzetta2008, Parker2009}.  In the canonical quantization of the matter fields, one typically requires    the resulting quantum theory to  preserve the symmetries of the classical theory. In the well-known case of flat spacetime  devoid of any external background field, this criterion turns out to select a unique quantization respecting Poincaré symmetry and the corresponding so-called Minkowski vacuum. However, when we introduce a time-dependent, external agent such as an electric field, time-translational invariance is broken. Then, the symmetry group of the classical theory is not large enough to completely determine a preferred quantum theory. Indeed, one encounters ambiguities in the choice of annihilation and creation operators, which lead to different quantum theories, each with their respective vacuum and particle notions. 
Even in vacuum, the action of the external field results in a nonvanishing spectral number $N(\tau)$ of particles produced at time $\tau$. This can be understood from the fact that the notion of particles has changed from the initial time to the time $\tau$. The precise number of particles, however, strongly depends on the specific vacuum states that we are comparing. In fact, for defining $N(\tau)$ and its time evolution, one needs to select one (global) notion of vacuum for each time $\tau$. This clearly poses questions about the physical interpretation of $N(\tau)$, and the discussion in the literature is still open~\cite{Ilderton2022,Alvarez2022,Dabrowski2014,Dabrowski2016,Yamada2021,Domcke2022,Diez2023}.

Recent works have experimentally implemented gravitational particle production in black hole \cite{Steinhauer2016, MunozDeNova2019} and cosmological \cite{Eckel2018, Wittemer2019, Steinhauer2022, Viermann2022} analog systems, where by means of two-point correlation functions of the density contrast, the number of produced particles after the expansion was measured. Motivated by the experimental accessibility of this quantity, we provide a way of understanding the physical meaning of the possible definitions of $N(\tau)$ in terms of the number of particles measured well after the time $\tau$ at which the external agent has been switched off.

We consider a simple setup in which an electric field is switched on smoothly from zero, so that there is no ambiguity in the choice of initial vacuum. In order to measure the actual number of particles at a certain time~$\tau_1$, one would need to instantaneously disconnect the external agent, here the background electric field, and measure afterward. However, instantaneous processes are unfeasible, and thus we cannot have experimental access to that magnitude. Instead, one possibility is to start switching the electric field off smoothly at that time, wait some time until the electric field is completely switched off, and finally measure. We denote this  outcome as $N^{\text{exp}}_{\tau_1}$. In order to measure the number of particles at a later time~$\tau_2$, we would need to repeat the experiment switching the external field off at that new instant. In this way, we would obtain a set of   measurement  results $N^{\text{exp}}_{\tau_1},   N^{\text{exp}}_{\tau_2},\cdots$, which tells us what is the number of particles measured in our experiment if we start to switch off the electric field at  $\tau_1, \tau_2, \cdots$.  Nevertheless, as we remarked before, this procedure and the results of the measurement will depend on how we switch the electric field off, which might be conditioned by the particular characteristics of the detector that we are using.

Here we propose to relate the different ways in which we can switch the electric field off and measure the number of particles on the one hand, with the theoretical ambiguities in the choice of the quantum vacuum on the other. For each measurement setup, leading to a family of results~$\{N_{\tau_i}^{\text{exp}}\}$, we can find among all possible quantizations at each $\tau_i$ a notion of vacuum such that $N(\tau_i)=N^{\text{exp}}_{\tau_i}$. The meaning of $N(\tau)$ becomes clear in this case: It is the resulting number of particles that would be measured, following our particular measurement process, if we switched off our experiment at time~$\tau$. Thus, canonical quantum ambiguities are inherently physical in the sense that they are intimately related to the infinitely many different ways of measuring.

\section{Preliminaries}

As a working case, let us consider a charged scalar field $\phi(t, x)$ in (1+1)-dimensional Minkowski spacetime in the presence of 
a spatially homogeneous time-dependent classical electric field, although our analysis can be extrapolated to higher dimensions, to similar particle creation scenarios due to an external time-dependent agent or to other matter fields (e.g., Dirac fields). Its dynamics is determined by the Klein-Gordon equation
\begin{equation}    \left[\left(\partial_{\mu}+iqA_{\mu}\right)\left(\partial^{\mu}+iqA^{\mu}\right)+m^2\right]\phi(t, x) = 0,
\label{eq:FieldEOM}
\end{equation}
where $m$ and $q$ are the mass and the charge of the field, respectively, 
and $A_{\mu}$ is the external electric potential. We choose the temporal gauge, namely $A_{\mu} = (0, A(t))$, and thus $E(t)=-\dot{A}(t)$, so that we explicitly preserve the spatial homogeneity of the electric field $E(t)$ in the quantization  of the scalar field. In this work we neglect backreaction of the quantum test fields. For supercritical electric field intensities and large particle densities, backreaction should be taken into account, for instance, using the generalized quantum Vlasov equation \cite{Alvarez2022}.

In order to construct a quantum field operator, we expand the  matter field in Fourier time-dependent modes~$\phi_k$ that verify decoupled harmonic oscillator equations
\begin{equation}
    \ddot{\phi}_k(t) + \Omega_k(t)^2 \phi_k(t)=0
    \label{eq:HOE}
\end{equation}
with time-dependent frequencies
\begin{equation}
    \Omega_k(t)^2 = k^2 + 2qA(t)k + q^2A(t)^2 + m^2.
\label{eq:Frequency}
\end{equation}
The expansion of each mode in terms of a basis of solutions $\varphi_k$ of Eq.~\eqref{eq:HOE}, 
\begin{equation}
\label{eq:hatphi}
     {\phi}_k(t)= {a}_k\varphi_k(t)+ {b}^{*}_k\varphi_k^*(t),
\end{equation}
determines the Fock quantization associated with this choice of basis, provided that the expansion coefficients are promoted to annihilation and creation operators $\hat{a}_k$ and~$\hat{b}^{\dagger}_k$, respectively. The  corresponding  quantum vacuum~$\ket{0}$ satisfies, by definition, $\hat{a}_k\ket{0}=\hat{b}_k\ket{0}=0$ for all $k$. Note that modes with different wave numbers $k$ are decoupled from one another. Thus, in the following we will drop the index $k$.

\begin{figure}[t!]
    \centering
    \includegraphics[width=0.42\textwidth]{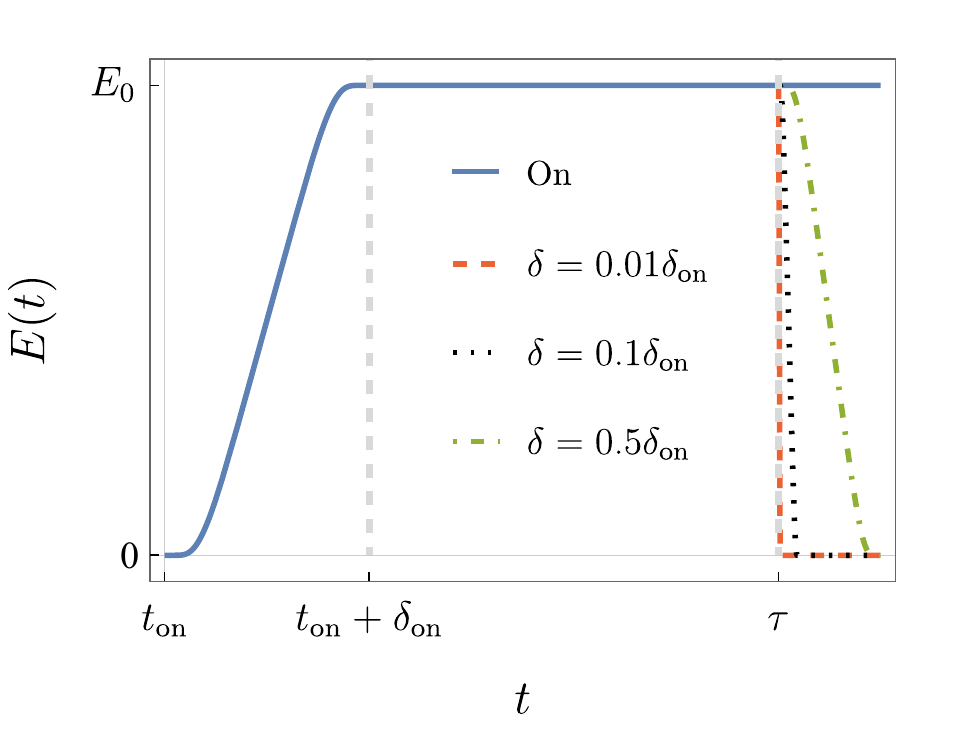}
    \caption{Time evolution of the electric field (solid line) with different switch off profiles (dashed/dotted lines) corresponding to different values of $\delta$, starting at $\tau$.}
    \label{fig:E(t)}
\end{figure}

We want to account for realistic (noninstantaneous) switch ons and switch offs of the electric field (see Fig.~\ref{fig:E(t)}). We start the experiment switching the electric field on at a time $t_{\text{on}}$, and after a time $\delta_{\text{on}}$ the electric field smoothly reaches the constant value $E_0$. Then, in order to measure the number of created particles at a given time $\tau$, we start switching it off at that time, and after a lapse~$\delta$, the electric field vanishes. This process is characterized by the properties of the experimental setup. 
The finite duration of the  on and off processes can be described by a $C^\infty$ step function $\Theta_{\sigma}(t)$ of width~$\sigma$ (i.e., $\delta_{\text{on}}$  and $\delta$, respectively, in our case at hand). The step Heaviside function corresponds to $\sigma=0$. The specific form of~$\Theta_{\sigma}(t)$ does not qualitatively affect the results. In our numerical computation, we use a $C^\infty$ regularization which interpolates between 0 and 1 in the interval $(-\sigma/2,\sigma/2)$ by means of the function \mbox{$\bm{(}1+\tanh \{\cot[\pi(1/2-t/\sigma)]\}\bm{)}/2$} and is constant outside said interval. In all figures we fix~$m=1$ and \mbox{$qE_0=1$}, so that the electric field reaches the critical Schwinger limit $m^2/q$~\cite{Schwinger1951,Yakimenko2019}. For lower strengths, the probability of producing a particle pair is negligible. In addition, we set $t_{\text{on}}=0$, and $A(t_{\text{on}})=0$, and times are given in units of the switch on duration $\delta_{\text{on}}$, while we parametrize different switch offs by varying $\delta$.

\section{Measured particle number}

Given a particular experimental setting, we can compute the asymptotic number of created particles $N^{\text{exp}}_{\tau}$ in the mode $k$ that would be measured by our detector when we start the switch off at the time $\tau$. Initially, when the electric field is not on yet, the matter field is in the Minkowski vacuum. This state is determined by the solution $\varphi^{\text{in}}$ to the mode equation~\eqref{eq:HOE} which behaves as a positive-frequency plane wave before~$t_{\text{on}}$. This is the only solution compatible with Poincaré symmetry in the asymptotic past. Analogously, there is another Minkowski out vacuum, associated with~$\varphi^{\text{out}}$, which is a positive-frequency plane wave at times after the electric field is completely switched off, namely, at $t_{\text{off}}=\tau+\delta$. However, because of the presence of the electric field, these two vacua are different. Indeed,~$N_{\tau}^{\text{exp}}$ measures how excited is the in vacuum with respect to the out vacuum. To obtain this quantity, we need to compare both solutions to Eq.~\eqref{eq:HOE} at the same time. Explicitly, we evolve the in solution from the initial time and compare it with $\varphi^{\text{out}}$, both evaluated at $t_{\text{off}}$. We can then calculate the asymptotic number of created particles in the mode $k$ according to the Bogoliubov formalism as
\begin{equation}
    N_{\tau}^{\text{exp}}=\left| \varphi^{\text{in}}(t_{\text{off}})\dot{\varphi}^{\text{out}}(t_{\text{off}})-\dot{\varphi}^{\text{in}}(t_{\text{off}})\varphi^{\text{out}}(t_{\text{off}}) \right|^2.
    \label{eq:numas}
\end{equation}

\begin{figure}[t!]
    \centering
    \includegraphics[width=0.495\textwidth]{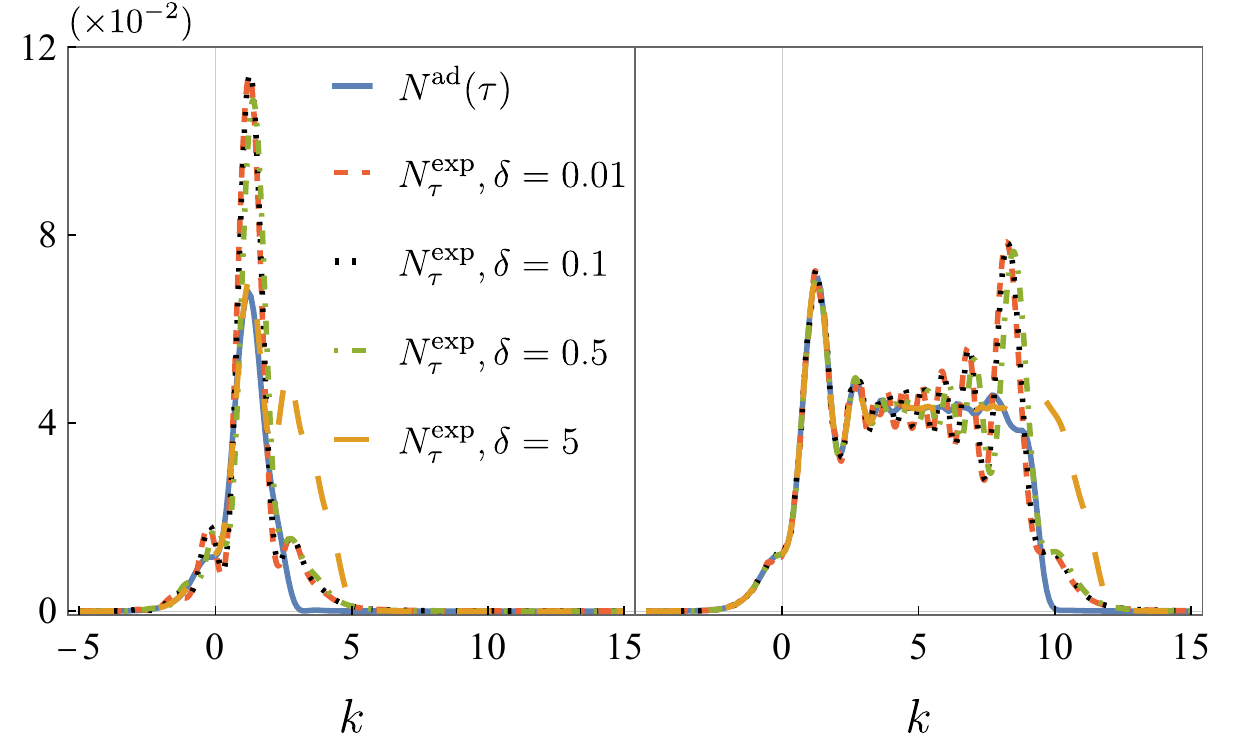}
    \caption{Spectra of the asymptotic number of created particles~$N_{\tau}^{\text{exp}}$ for different switch offs, starting at times $\tau=3$ (left) and $\tau=10$ (right), for different switch off durations $\delta$ (dashed lines). We also represent the computed number of created particles $N^\text{ad}(\tau)$ in the zeroth-order adiabatic vacuum at those times (solid line).}
    \label{fig:numk}
\end{figure}

It is important to remark that for a different measurement process starting at the same $\tau$, here characterized by another duration of the switch off $\delta$, the outcome $N_{\tau}^{\text{exp}}$ will change. In Fig.~\ref{fig:numk}, we show the spectra of asymptotically produced particles $N_{\tau}^{\text{exp}}$ for $\tau=3$ and $\tau=10$ for different switch off durations $\delta$. Observe that the slower the switch offs, the longer the electric field can accelerate particles, and thus modes with larger $k$ become excited. This behavior is in agreement with that of Ref. \cite{Adorno2018}, where they thoroughly analyze the role played by $\delta$ and~$\tau$ in particle production for a similar profile of the electric field. Note that the oscillations in the spectral distribution have already been observed in analog experiments by means of two-point correlation functions~\cite{Steinhauer2022,Viermann2022}.

\section{Theoretical particle number}

On the other hand, one may be interested in theoretically computing the number of particles that have been created from an initial time~$t_{\text{on}}$ to some time $\tau$, without relying on any measurement. This question is, however, much more subtle. One would need to choose a particular solution~$\varphi^{\tau}$ to Eq.~\eqref{eq:HOE} by imposing initial conditions at time~$\tau$, i.e., $(\varphi^{\tau}(\tau),\dot{\varphi}^{\tau}(\tau))$. Contrary to what happens for the out-solution $\varphi^{\text{out}}$, there is an ambiguity in the selection of $\varphi^{\tau}$. Indeed, when the electric field is on, the frequency \eqref{eq:Frequency} is not constant and the physical criterion of preservation of the classical symmetries in the quantum theory is not strong enough to fix a unique vacuum. The spectral number of created particles between $t_{\text{on}}$ and $\tau$ is defined by
\begin{equation}
    N(\tau)=\left| \varphi^{\text{in}}(\tau)\dot{\varphi}^{\tau}(\tau)-\dot{\varphi}^{\text{in}}(\tau)\varphi^{\tau}(\tau) \right|^2.
    \label{eq:numt}
\end{equation}
In contrast with the measured value $N_{\tau}^{\text{exp}}$, which is unique given a particular switch off of the electric field starting at a certain time $\tau$, $N(\tau)$ strongly depends on the choice of vacuum at that time. This is precisely the reason why the interpretation of this magnitude is not clear in the literature yet \cite{Ilderton2022,Alvarez2022,Dabrowski2014,Dabrowski2016,Yamada2021}.

As a well-known example, the solid line in Fig.~\ref{fig:numk} shows the computed number of created particles $N^\text{ad}(\tau)$ for the zeroth-order adiabatic initial conditions~\cite{Birrell1982}
\begin{equation}
\begin{split}
\label{eq:ad0}
    \varphi^{\text{ad}}(\tau)&=\frac{1}{\sqrt{2\Omega_k(\tau)}}, \\ 
    \dot{\varphi}^{\text{ad}}(\tau)&=-i\sqrt{\frac{\Omega_k(\tau)}{2}}-\frac{\dot{\Omega}_k(\tau)}{2\sqrt{2}\Omega_k(\tau)}.
\end{split}
\end{equation}
Note the differences between this curve and the corresponding to very fast switch offs. We will comment on this later on.

\section{Relation between measured and theoretical particle numbers}

Each measurement procedure selects a particular vacuum for which the theoretical number of particles $N(\tau)$ has a well-defined physical meaning. Indeed, among all possibilities for choosing a particular mode $\varphi^{\tau}$ in Eq.~\eqref{eq:numt}, there is one for which the number of predicted particles at time $\tau$ coincides with the outcome $N^{\text{exp}}_{\tau}$ that a particular measurement device would yield. More explicitly, for each time $\tau$, we choose the mode~$\varphi^{\tau}$ as the out vacuum $\varphi^{\text{out}}$  associated with the switch off starting at that time. In fact, replacing $t_{\text{off}}$ by $\tau$ in Eq.~\eqref{eq:numas} does not change the resulting $N^{\text{exp}}_{\tau}$, since this magnitude does not depend on the instant at which it is evaluated. Consequently, the choice $\varphi^{\tau}=\varphi^{\text{out}}$ makes Eq.~\eqref{eq:numt} equal to Eq.~\eqref{eq:numas}. The set of modes $\{\varphi^{\text{exp}, \tau}\}$ and corresponding vacua $\{\ket{0}^{\text{exp},\tau}\}$ defined in this way, one for each $\tau$, allows one to construct $N^{\text{exp}}(\tau)$, viewed as function of the time at which we start switching the electric field off. At each time $\tau$, the in vacuum is an excited state with respect to the vacuum $\ket{0}^{\text{exp},\tau}$: Its excitations correspond precisely to the quanta that would be measured by our detector. Note that $\tau$ denotes the time at which we want to calculate the particles produced by the electric field and not the starting point of a programmed switch off.

This prescription defines a family of physical vacua: those for which there exists a switch off giving the same particle number as the one predicted by the vacuum. Furthermore, all these vacua unitarily implement the dynamics, as they are associated with a finite number of particles by construction, thus verifying Shale's theorem~\cite{Shale1962}.

\begin{figure}[t!]
    \centering
    \includegraphics[width=0.49\textwidth]{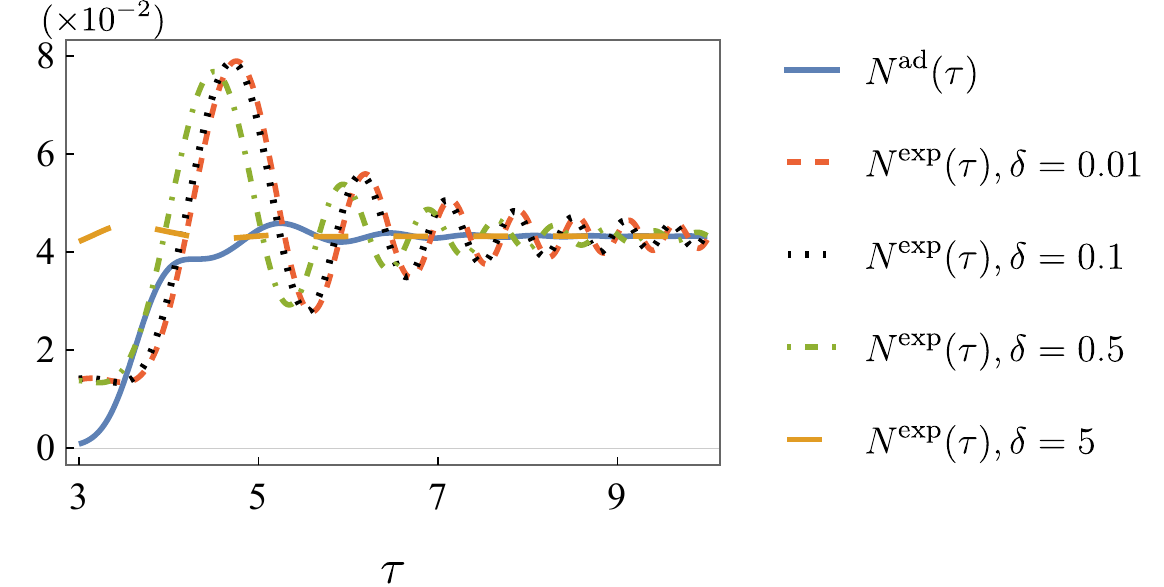}
    \caption{Evolution of the number of created particles $N^{\text{exp}}(\tau)$ with $k=3$ for different switch off durations~$\delta$. The solid line corresponds to the zeroth-order adiabatic prescription.}
    \label{fig:numt}
\end{figure}

In our simple setup, the measurement device is characterized by $\delta$, and for each of its values we have a different set of modes $\{\varphi^{\text{exp}, \tau}\}$ and hence different notions of~$N^{\text{exp}}(\tau)$. This is illustrated in Fig.~\ref{fig:numt}, where one can see the time evolution of $N^{\text{exp}}(\tau)$ for $k=3$, for different durations of the switch off~$\delta$. For each time~$\tau$, we compute the asymptotic number of particles $N_{\tau}^{\text{exp}}$ when we start switching the electric field off at $\tau$. The observed oscillations in~$\tau$ were already present in Refs. \cite{Kluger1998,Schmidt1998,Dabrowski2014,Dabrowski2016}, but now we can provide them with a full physical meaning, as they follow from a measurement-based notion of particle. Moreover, recent works try to implement experimental setups that make use of this behavior to enhance particle production (see the recent study~\cite{Aleksandrov2022b} or other references on the dynamically assisted Schwinger effect~\cite{Schutzhold2008,Bulanov2010}). In Fig. \ref{fig:numt} we also show the time evolution of the theoretical particle number when we choose zeroth-order adiabatic vacua at each time $\tau$.

The amplitudes of the fluctuations are smaller as we increase the value of~$\delta$. Note that as $\tau$ increases, these amplitudes decrease and the number of particles become more independent of $\delta$. This result is compatible with~\cite{Adorno2018}, where it was proved that for a sufficiently large time $\tau$ (larger than the ones considered here) the switch on and off effects only affect as next-to-leading corrections to the contribution to the constant part of the electric field.

As an aside, for each measurement process we could have reassigned the asymptotic outcome $N_{\tau}^{\text{exp}}$ to any other time different from the time at which the switch off starts (e.g., results in~\cite{Ilderton2022} are obtained by taking $\tau+\delta/2$ as the reference time instead). However, note that this would lead to a simple relabeling of the $\tau$~axis in Fig.~\ref{fig:numt}, shifting each curve proportionally to its value of~$\delta$.

Finally, one may wonder whether it is possible to find, given a choice of vacuum, a particular switch off starting at $\tau$ such that the measurement of the associated asymptotic number of particles~$N_{\tau}^{\text{exp}}$ coincides with the value of $N(\tau)$, computed using said vacuum prescription. This requirement does not unequivocally determine the time evolution of the in mode $\varphi^{\text{in}}$ after time $\tau$. Therefore, each function $\varphi^{\text{in}}$ compatible with the previous condition would lead to a different mode equation. However, it is definitely nontrivial that one can find a time-dependent frequency~$\Omega(t)$ of the form of Eq.~\eqref{eq:Frequency} fulfilling this requirement for all values of $k$.

\section{Interpretation of usual vacuum prescriptions}

We now illustrate the application of our operational notion of particles by interpreting two usual notions of vacuum in terms of measurements. First, we consider the so-called instantaneous lowest-energy state (ILES), which is used in many references, especially in those studying the quantum Vlasov equation~\cite{Kluger1998,Schmidt1998,Aleksandrov2022,Mottola2014,Ruffini2010,Roberts2000,Dunne2009,Dumlu2011,Hebenstreit2009}. It minimizes the energy per mode at a particular instant of time~$\tau$ and corresponds to the solution to Eq.~\eqref{eq:HOE} with initial conditions
\begin{equation}
\label{eq:ILES}
    \varphi^{\text{ILES}}(\tau)=\frac{1}{\sqrt{2\Omega_k(\tau)}}, \ \dot{\varphi}^{\text{ILES}}(\tau)=-i\sqrt{\frac{\Omega_k(\tau)}{2}}.
\end{equation}
In~\cite{Ilderton2022}, it was proved that the theoretical particle number calculated using this vacuum coincides with the asymptotic particle number measured in the unfeasible situation in which the electric field is instantaneously switched off at $\tau$, i.e., $\delta=0$. Indeed, this setting can be easily implemented in the electric potential with a continuous but nondifferentiable step function at $\tau$. The most regular solution to Eq.~\eqref{eq:HOE} has a continuous but nondifferentiable second derivative and corresponds precisely to the instantaneous lowest-energy state~\eqref{eq:ILES}. In addition, in agreement with Ref.~\cite{Ilderton2022}, the particle number in the instantaneous case $\delta=0$ coincides with the limit $\delta \to 0$.

Another vacuum prescription that is commonly used is precisely the one defined in Eq. \eqref{eq:ad0}, which is given by a WKB approximation of higher order than the previous instantaneous lowest-energy state defined in~\eqref{eq:ILES}. This vacuum is usually called the zeroth-order adiabatic vacuum, although this name is sometimes used for the prescription~\eqref{eq:ILES}. We infer from Figs.~\ref{fig:numk} and~\ref{fig:numt} that the particle number spectrum of the adiabatic vacuum deviates from that of an arbitrarily fast switch off. Indeed, in the latter, there appear fluctuations with larger amplitudes both in the spectrum and in the time evolution.

\section{Conclusions}

Quantum vacuum ambiguities are inherent to quantum field theory in the presence of an external, time-dependent agent. In this work, we show that knowing the particularities of how we measure the particle number allows us to identify a particular quantum vacuum with clear physical meaning: Its associated notion of particle is that which would be measured by our detector in a potential experiment.

This operational procedure can be used to interpret the notion of particle associated with usual vacuum prescriptions. This is the case, for example, of the instantaneous lowest-energy state at a certain time, which provides the same particle number as an instantaneous switch off at that time or of the zeroth-order adiabatic vacuum, which departs from this behavior.

In conclusion, we select a family of vacua---those related to a realistic switch off of the external agent---that are physical in the sense that they accommodate information about real outcomes. These vacua are intrinsically well-behaved as they allow for a unitary implementation of the dynamics.

\acknowledgments

This work is partially supported by the MICINN (Spain) projects PID2019-107394GB-I00/AEI/10.13039/501100011033 (AEI/FEDER, UE), PID2019-108655GB-I00 (AEI), PID2020-118159GB-C44, and PID2022-139841NB-I00, the EU STRONG-2020 project (Grant No. 824093), COST (European Cooperation in Science and Technology) Actions CA21106 and CA21136, and STMS Grant from COST Action CA16108. Á.P.-L. acknowledges financial support from the MIU (Ministerio de Universidades, Spain) fellowship FPU20/05603. J.M.S.V. acknowledges the support of the Spanish Agencia Estatal de Investigaci\'on through the grant “IFT Centro de Excelencia Severo Ochoa CEX2020-001007-S."


\bibliography{Bibliography.bib}

\end{document}